\documentclass{article}
\usepackage{cite}
\usepackage{graphicx}
\usepackage{dcolumn}

\begin{document}

\title{Comment on: ``On the effects of the Lorentz symmetry violation yielded by a
tensor field on the interaction of a scalar particle and a Coulomb-type
field'' Ann. Phys. 399 (2018) 117-123}
\author{Paolo Amore\thanks{%
e--mail: paolo@ucol.mx} \\
Facultad de Ciencias, CUICBAS, Universidad de Colima,\\
Bernal D\'{\i}az del Castillo 340, Colima, Colima,Mexico \\
and \\
Francisco M. Fern\'andez\thanks{%
e--mail: fernande@quimica.unlp.edu.ar} \\
INIFTA, Divisi\'{o}n Qu\'{\i}mica Te\'{o}rica,\\
Blvd. 113 y 64 (S/N), Sucursal 4, Casilla de Correo 16,\\
1900 La Plata, Argentina}
\maketitle

\begin{abstract}
We analyze the eigenvalues and eigenfunctions stemming from a recent study
of the interaction of a scalar particle with a Coulomb potential in the
presence of a background of the violation of the Lorentz symmetry
established by a tensor field. We show, beyond any doubt, that the physical
conclusions drawn by the authors from a truncation of a power series, coming
from the application of the Frobenius method, are meaningless and
nonsensical.
\end{abstract}

In a recent paper Vit\'{o}ria et al\cite{VBB18} analyze the interaction of a
scalar particle with a Coulomb-type potential in the presence of a
background of the violation of the Lorentz symmetry established by a tensor
field. The equation proposed by the authors is separable in cylindrical
coordinates and the radial part is a solution to an eigenvalue equation with
centrifugal-like ($r^{-2}$), Coulomb ($r^{-1}$) and harmonic ($r^{2}$)
terms. The application of the Frobenius method leads to a three-term
recurrence relation for the expansion coefficients and the authors force a
truncation in order to obtain polynomial solutions. In this way they obtain
analytical expressions for the energies of the system and conclude that
there are \textit{permitted} values of a parameter that characterizes the
magnetic field. The purpose of this Comment is the analysis of the effect of
the truncation approach on the physical conclusions drawn by the authors.

The starting point of present discussion is the eigenvalue equation for the
radial part of the Schr\"{o}dinger equation
\begin{eqnarray}
&&F^{\prime \prime }(x)+\frac{1}{x}F^{\prime }(x)-\frac{\gamma ^{2}}{x^{2}}%
F(x)+\frac{\theta }{x}F(x)-x^{2}F(x)+WF(x)=0,  \nonumber \\
&&W=\frac{\beta }{\tau },\;\beta =\mathcal{E}^{2}-m^{2}-p_{z}^{2},\;\gamma
^{2}=l^{2}-\alpha ^{2},\;\tau ^{2}=\frac{1}{2}gb\chi ^{2},  \nonumber \\
&&\theta =\frac{2\alpha \mathcal{E}}{\sqrt{\tau }}  \label{eq:eig_eq}
\end{eqnarray}
where $l=0,\pm 1,\pm 2,\ldots $ is the rotational quantum number (restricted
to $l^{2}\geq \alpha ^{2}$), $m$ the mass of the particle, $\alpha $ the
strength of a Coulomb-type potential, $\mathcal{E}$ the energy, $b=-\left(
K_{HB}\right) _{zz}>0$, $\chi $ comes from a magnetic field and $g$ is a
constant. The constant $-\infty <p_{z}<\infty $ is the quantum number for
the free motion along the $z$ direction. The authors simply set $\hbar =1$, $%
c=1$ though there are well known procedures for obtaining suitable
dimensionless equations in a clearer and more rigorous way\cite{F20}. In
what follows we focus on the discrete values of $W$ that one obtains from
the bound-state solutions of equation (\ref{eq:eig_eq}) that satisfy
\begin{equation}
\int_{0}^{\infty }\left| F(x)\right| ^{2}x\,dx<\infty .
\label{eq:bound_states}
\end{equation}
Notice that we have bound states for all $-\infty <\theta <\infty $ and that
the eigenvalues $W$ satisfy
\begin{equation}
\frac{\partial W}{\partial \theta }=-\left\langle \frac{1}{x}\right\rangle
<0,  \label{eq:HFT}
\end{equation}
according to the Hellmann-Feynman theorem\cite{F39}.

The eigenvalue equation (\ref{eq:eig_eq}) is an example of conditionally
solvable (or quasi-exactly solvable) problems that have been widely studied
by several authors and exhibit a hidden algebraic structure (see, for
example, \cite{T16} and references therein).

In order to solve the eigenvalue equation (\ref{eq:eig_eq}) the authors
proposed the ansatz
\begin{equation}
F(x)=x^{s}\exp \left( -\frac{x^{2}}{2}\right) P(x),\;P(x)=\sum_{j=0}^{\infty
}a_{j}x^{j},\;s=|\gamma |,  \label{eq:ansatz}
\end{equation}
and derived the three-term recurrence relation
\begin{eqnarray}
a_{j+2} &=&-\frac{\theta }{\left( j+2\right) \left[ j+2\left( s+1\right)
\right] }a_{j+1}+\frac{2j+2s-W+2}{\left( j+2\right) \left[ j+2\left(
s+1\right) \right] }a_{j},\;  \nonumber \\
j &=&-1,0,\ldots ,\;a_{-1}=0,\;a_{0}=1.  \label{eq:TTRR}
\end{eqnarray}

If the truncation condition $a_{n+1}=a_{n+2}=0$ has physically acceptable
solutions then one obtains some exact eigenvalues and eigenfunctions. The
reason is that $a_{j}=0$ for all $j>n$ and the factor $P(x)$ in equation (%
\ref{eq:ansatz}) reduces to a polynomial of degree $n$. This truncation
condition is equivalent to $W_{s}^{(n)}=2(n+s+1)$ and $a_{n+1}=0$. The
latter equation is a polynomial function of $\theta $ of degree $n+1$ and it
can be proved that all the roots $\theta _{s}^{(n,i)}$, $i=1,2,\ldots ,n+1$,
$\theta _{s}^{(n,i)}<\theta _{s}^{(n,i+1)}$, are real\cite{CDW00,AF20}. If $%
V(\theta ,x)=-\theta /x+x^{2}$ denotes the parameter-dependent potential for
the model discussed here, then it is clear that the truncation condition
produces an eigenvalue $W_{s}^{(n)}$ that is common to $n+1$ different
potential-energy functions $V_{s}^{(n,i)}(x)=V\left( \theta
_{s}^{(n,i)},x\right) $. Notice that in this analysis we have deliberately
omitted part of the interaction that has been absorbed into $\gamma $ (or $s$%
) because it is not affected by the truncation approach. It is also worth
noticing that the truncation condition only yields \textit{some particular}
eigenvalues and eigenfunctions because not all the solutions $F(x)$ of (\ref
{eq:eig_eq}) satisfying equation (\ref{eq:bound_states}) have polynomial
factors $P(x)$. From now on we will refer to them as follows
\begin{equation}
F_{s}^{(n,i)}(x)=x^{s}P_{s}^{(n,i)}(x)\exp \left( -\frac{x^{2}}{2}\right)
,\;P_{s}^{(n,i)}(x)=\sum_{j=0}^{n}a_{j,s}^{(n,i)}x^{j}.
\label{eq:f^(n,i)(y)}
\end{equation}
We want to stress that the $n+1$ eigenfunctions $F_{s}^{(n,i)}(x)$, $%
i=1,2,\ldots ,n+1$ share the \textit{same} eigenvalue $W_{s}^{(n)}$, a point
that was not taken into account by Vit\'{o}ria et al\cite{VBB18} and that is
of utmost relevance, as shown below.

Let us consider the first cases as illustrative examples. When $n=0$ we have
$W_{s}^{(0)}=2(s+1)$, $\theta _{s}^{(0)}=0$ and the eigenfunction $%
F_{s}^{(0)}(x)$ has no nodes. We may consider this case trivial because the
problem reduces to the exactly solvable harmonic oscillator. Probably for
this reason it was not explicitly considered by Vit\'{o}ria et al\cite{VBB18}%
.

When $n=1$ there are two roots $\theta _{s}^{(1,1)}=-\sqrt{4s+2}$ and $%
\theta _{s}^{(1,2)}=\sqrt{4s+2}$ and the corresponding non-zero coefficients
are
\begin{equation}
a_{1,s}^{(1,1)}=\frac{\sqrt{2}}{\sqrt{2s+1}},\;a_{1,s}^{(1,2)}=-\frac{\sqrt{2%
}}{\sqrt{2s+1}},  \label{eq:a^(1,i)_1}
\end{equation}
respectively. We appreciate that the eigenfunction $F_{s}^{(1,1)}(x)$ is
nodeless and $F_{s}^{(1,2)}(x)$ has one node and that both corresponds to
the \textit{same} eigenvalue $W_{s}^{(1)}$.

When $n=2$ the results are
\begin{eqnarray}
\theta _{s}^{(2,1)} &=&-2\sqrt{4s+3},\;a_{1,s}^{(2,1)}=\frac{2\sqrt{4s+3}}{%
2s+1},\;a_{2,s}^{(2,1)}=\frac{2}{2s+1},  \nonumber \\
\theta _{s}^{(2,2)} &=&0,\;a_{1,s}^{(2,2)}=0,\;a_{2,s}^{(2,2)}=-\frac{1}{s+1}%
,  \nonumber \\
\theta _{s}^{(2,3)} &=&2\sqrt{4s+3},\;a_{1,s}^{(2,3)}=-\frac{2\sqrt{4s+3}}{%
2s+1},\;a_{2,s}^{(2,3)}=\frac{2}{2s+1}.  \label{eq:alpha,a_n=2}
\end{eqnarray}
Notice that $F_{s}^{(2,1)}(x)$, $F_{s}^{(2,2)}(x)$ and $F_{s}^{(2,3)}(x)$
have zero, one and two nodes, respectively, in the interval $0<x<\infty $
and that the three eigenfunctions correspond to the \textit{same} eigenvalue
$W_{s}^{(2)}$.

From the truncation condition the authors derived
\begin{equation}
\mathcal{E}_{n,l,p_{z}}^{2}=m^{2}+p_{z}^{2}+2\tau \left( n+|\gamma
|+1\right) ,  \label{eq:E^2_VBB}
\end{equation}
as well as expressions for $\tau _{n,l,p_{z}}$ and $\chi _{n,l,p_{z}}$, $%
n=1,2$. They concluded that there are \textit{permitted} values of $\chi $
that characterize the magnetic field. Since there are square-integrable
solutions to the eigenvalue equation (\ref{eq:eig_eq}) for all values of $%
\theta $ it is clear that such particular values of $\chi $ are just an
artifact of the truncation method that yields particular solutions to the
eigenvalue equation with polynomial factors $P_{s}^{(n,i)}(x)$. Besides, the
allowed energies associated to the nodes $n=1$ and $n=2$ obtained by the
authors have no physical meaning because they stem from different potentials
$V_{s}^{(n,i)}(x)$. In what follows we discuss this point with more detail.

In order to make present discussion clear we write the \textit{actual}
eigenvalues of equation (\ref{eq:eig_eq}) as $W_{j,s}(\theta )$, $%
j=0,1,\ldots $, $W_{j,s}<W_{j+1,s}$. Given that there are square-integrable
solutions for all $-\infty <\theta <\infty $, as indicated above, each
eigenvalue can be considered to be a curve $W_{j,s}(\theta )$ in the $\left(
\theta ,W\right) $ plane. Therefore, the correct energies of the system
should be
\begin{equation}
\mathcal{E}_{j,l,p_{z}}^{2}=m^{2}+p_{z}^{2}+\tau W_{j,s}.
\label{eq:E^2_present}
\end{equation}
Since the eigenvalue equation (\ref{eq:eig_eq}) is not exactly solvable,
except for some particular values of $\theta $, we should resort to an
approximate method in order to obtain the eigenvalues and eigenfunctions
that are not given by the truncation condition. Here, we apply the well
known Rayleigh-Ritz variational method that yields upper bounds to all the
eigenvalues and choose the non-orthogonal basis set $\left\{ x^{s+j}\exp
\left( -\frac{x^{2}}{2}\right) ,\;j=0,1,\ldots \right\} $.

We arbitrarily choose $s=0$ as a first illustrative example in order to
facilitate the calculations. When $\theta =\theta _{0}^{(1,1)}=-\sqrt{2}$
the first four eigenvalues are $W_{0,0}=W_{0}^{(1)}=4$, $W_{1,0}=7.693978891$%
, $W_{2,0}=11.50604238$, $W_{3,0}=15.37592718$; on the other hand, when $%
\theta =\theta _{0}^{(1,2)}=\sqrt{2}$ we have $W_{0,0}=-1.459587134$, $%
W_{1,0}=W_{0}^{(1)}=4$, $W_{2,0}=8.344349427$, $W_{3,0}=12.53290130$. Notice
that the truncation condition yields only the ground state for the former
model and the first excited state for the latter, missing all the other
eigenvalues for each model potential.

As a second example we choose $s=1$, again to facilitate the calculations.
When $\theta =\theta _{1}^{(1,1)}=-\sqrt{6}$ the first four eigenvalues are $%
W_{0,0}=W_{1}^{(1)}=6$, $W_{1,1}=9.805784090$, $W_{2,1}=13.66928892$, $%
W_{3,1}=17.56601881$; on the other hand, when $\theta =\theta _{1}^{(1,2)}=%
\sqrt{6}$ we have $W_{0,1}=1.600357154$, $W_{1,1}=W_{1}^{(1)}=6$, $%
W_{2,1}=10.21072810$, $W_{3,1}=14.35078474$. Notice that the truncation
condition yields only the lowest state for the former model and the
second-lowest one for the latter, missing all the other eigenvalues for each
model potential.

In order to convince the reader about the accuracy of the variational
method, tables \ref{tab:theta1} and \ref{tab:theta2} show how the
approximate eigenvalues given by this approach converge from above towards
the exact eigenvalues of equation (\ref{eq:eig_eq}) as the number $N$ of
functions in the expansion increases. We appreciate that the variational
method yields the exact eigenvalue $W_{1}^{(1)}$ for all $N$ because the
corresponding eigenfunction is, in this case, a linear combination of only
two basis functions.

\begin{figure}[tbp]
\begin{center}
\includegraphics[width=9cm]{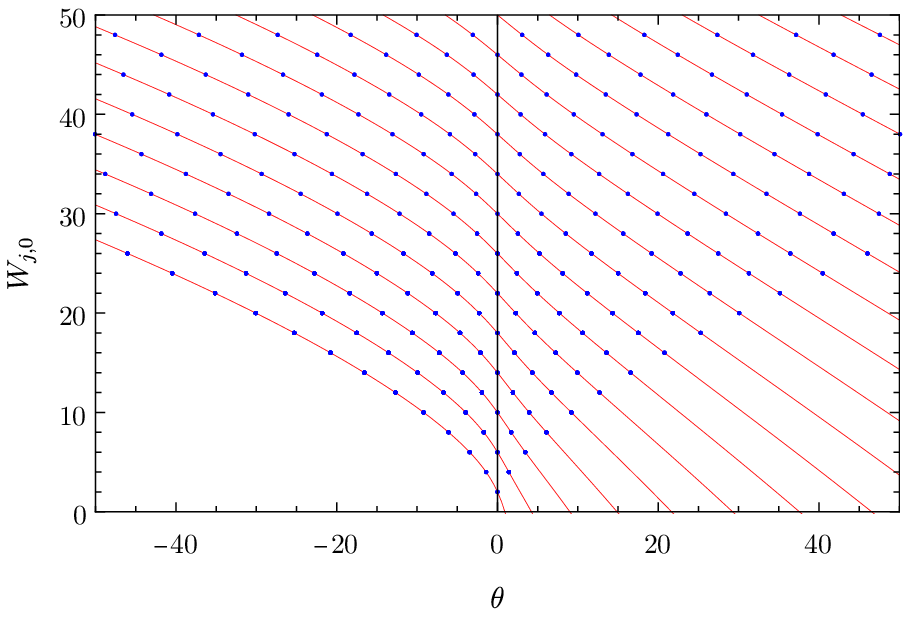} \includegraphics[width=9cm]{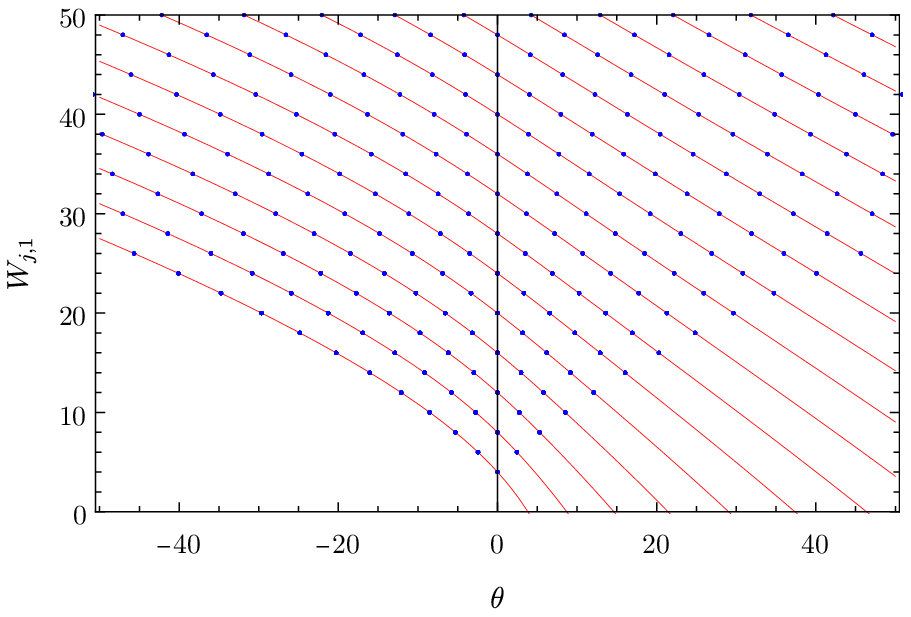}
\end{center}
\caption{Eigenvalues $W_{j,0}$ (upper panel) and $W_{j,1}$ (lower panel)
obtained from the truncation condition (blue points) and from the
variational method (red lines)}
\label{Fig:Wn}
\end{figure}

From the analysis above one may draw the wrong conclusion that the
truncation condition is utterly useless; however, it has been shown that one
can extract valuable information about the spectrum of conditionally
solvable models if one arranges and connects the roots $W_{s}^{(n)}$ properly%
\cite{CDW00,AF20}. From the analysis outlined above we conclude that $\left(
\theta _{s}^{(n,i)},W_{s}^{(n)}\right) $ is a point on the curve $%
W_{i-1,s}(\theta )$, $i=1,2,\ldots ,n+1$, so that we can easily construct
some parts of such spectral curves. For example, Figure~\ref{Fig:Wn} shows
several eigenvalues $W_{0}^{(n)}$ and $W_{1}^{(n)}$ given by the truncation
condition (blue points) and red lines representing the variational
calculations. Notice that the continuous variational curves $W_{j,s}(\theta
) $ already connect the points $W_{s}^{(n)}$ corresponding to the truncation
condition. In other words, the variational method yields \textit{all} the
eigenvalues $W_{j,s}(\theta )$ for any value of $\theta $ while the
truncation results $W_{s}^{(n)}$ are just \textit{some particular} points on
the curves. Besides, it is clear that the variational curves $W_{j,s}(\theta
)$ have negative slopes as predicted by the Hellmann-Feynman theorem (\ref
{eq:HFT}). We clearly see that the \textit{allowed} energies reported by
Vit\'{o}ria et al\cite{VBB18} have no physical meaning because they
correspond to many different problems instead of just one. In addition to
it, the occurrence of discrete \textit{permitted} values of the magnetic
field parameter $\chi $ is a mere consequence of selecting particular points
$\left( \theta _{s}^{(n,i)},W_{s}^{(n)}\right) $ on the curves $%
W_{j,s}(\theta )$. It should be clear from present analysis that such points
(by themselves) do not exhibit any physical meaning. Notice that the
truncation method only yields the exact result in the trivial case $\theta
=0 $. This fact is already discussed in many textbooks of quantum mechanics
where it is shown that the coefficients of the power series expansions of
the solutions to the exactly solvable quantum-mechanical models, like the
harmonic oscillator, hydrogen atom, Morse oscillator, etc., satisfy two-term
recurrence relations and not three-term ones like the quasi-exactly solvable
problems\cite{CDW00,AF20}.

In two earlier papers on this journal Bakke\cite{B14} and Bakke and Furtado%
\cite{BF15} discussed physical systems with different interactions, arrived
at the same eigenvalue equation, applied the same approach and,
consequently, draw somewhat similar wrong physical conclusions.

\section*{Acknowledgements}

The research of P.A. was supported by Sistema Nacional de
Investigadores (M\'exico).

\section*{Addendum}

According to the reviewer: ``Another point to be observed is the
dependence of the Rayleigh-Ritz variational method on the choice
of the wave function. Despite not being mentioned by the authors
of this comment, the wave function used in the Rayleigh-Ritz
variational method is obtained from the asymptotic analysis made
by Vit\'{o}ria et al. If one uses another wave function that
differs from the wave function obtained from the asymptotic
analysis made by Vit\'{o}ria et al, therefore, the results will be
different. In addition, no mathematical proof has been shown in
this comment that clarifies the relation of the approximate
solutions to the biconfluent Heun equation.''

This comment is surprising. In order to apply the Ritz variational
method it is mandatory that the basis functions satisfy the
correct boundary conditions at $x=0$ and $x\rightarrow \infty $.
Therefore, we have chosen the simplest basis set that satisfy such
boundary conditions. The  set of Gaussian functions chosen here is
complete and, for this reason it should give the actual
eigenvalues of the problem at hand. This fact is clearly revealed
in the convergence of the approximate eigenvalues shown in Tables
\ref{tab:theta1} and \ref{tab:theta2}. The Ritz variational method
is well known and and has been widely used for the study of many
quantum-mechanical problems.

\begin{table}[tbp]
\caption{Eigenvalues $W_{j,0}$ for $\gamma=0$ and $\theta=-\protect\sqrt{2}$}
\label{tab:theta1}
\begin{center}
\par
\begin{tabular}{D{.}{.}{3}D{.}{.}{11}D{.}{.}{11}D{.}{.}{11}D{.}{.}{11}}
\hline \multicolumn{1}{c}{$N$}& \multicolumn{1}{c}{$W_{0 0}$} &
\multicolumn{1}{c}{$W_{1 0}$} & \multicolumn{1}{c}{$W_{2 0}$}  & \multicolumn{1}{c}{$W_{3 0}$}\\
\hline

 2  &  4.000000000  &   10.49997602  &                &                \\
 3  &  4.000000000  &   7.751061995  &   19.88102859  &                \\
 4  &  4.000000000  &   7.694010921  &   11.97562584  &   33.92039998  \\
 5  &  4.000000000  &   7.693979367  &   11.51212379  &   17.05520450  \\
 6  &  4.000000000  &   7.693978905  &   11.50604696  &   15.46896992  \\
 7  &  4.000000000  &   7.693978892  &   11.50604243  &   15.37652840  \\
 8  &  4.000000000  &   7.693978891  &   11.50604238  &   15.37592761  \\
 9  &  4.000000000  &   7.693978891  &   11.50604238  &   15.37592718  \\
10  &  4.000000000  &   7.693978891  &   11.50604238  &   15.37592718  \\
\end{tabular}
\par
\end{center}
\end{table}

\begin{table}[tbp]
\caption{Eigenvalues $W_{j,0}$ for $\gamma=0$ and $\theta=\protect\sqrt{2}$}
\label{tab:theta2}
\begin{center}
\par
\begin{tabular}{D{.}{.}{3}D{.}{.}{11}D{.}{.}{11}D{.}{.}{11}D{.}{.}{11}}
\hline \multicolumn{1}{c}{$N$}& \multicolumn{1}{c}{$W_{0 0}$} &
\multicolumn{1}{c}{$W_{1 0}$} & \multicolumn{1}{c}{$W_{2 0}$}  & \multicolumn{1}{c}{$W_{3 0}$}\\
\hline

 2&  -1.180391283 &  4.000000000 &              &               \\
 3&  -1.401182256 &  4.000000000 &  9.284143096 &               \\
 4&  -1.449885589 &  4.000000000 &  8.345259771 &  17.66452696  \\
 5&  -1.458156835 &  4.000000000 &  8.344361267 &  12.69095166  \\
 6&  -1.459389344 &  4.000000000 &  8.344349784 &  12.53313315  \\
 7&  -1.459560848 &  4.000000000 &  8.344349442 &  12.53290257  \\
 8&  -1.459583736 &  4.000000000 &  8.344349427 &  12.53290132  \\
 9&  -1.459586704 &  4.000000000 &  8.344349427 &  12.53290130  \\
10&  -1.459587081 &  4.000000000 &  8.344349427 &  12.53290130  \\
11 &  -1.459587128 & 4.000000000 & 8.344349427 & 12.53290130  \\
12 &  -1.459587134 & 4.000000000 & 8.344349427 & 12.53290130  \\
13 &  -1.459587134 & 4.000000000 & 8.344349427 & 12.53290130  \\
\end{tabular}
\par
\end{center}
\end{table}

\end{document}